\documentclass[sn-mathphys-num]{sn-jnl}

\usepackage{graphicx}
\usepackage{multirow}
\usepackage{amsmath,amssymb,amsfonts}
\usepackage{amsthm}
\usepackage{mathrsfs}
\usepackage[title]{appendix}
\usepackage{xcolor}
\usepackage{textcomp}
\usepackage{manyfoot}
\usepackage{booktabs}
\usepackage{algorithm}
\usepackage{algpseudocode}
\usepackage{listings}
\usepackage{caption}
\usepackage{subcaption}
\usepackage{url}
\usepackage{tikz}
\usetikzlibrary{external}
\usetikzlibrary{shapes.geometric, arrows, arrows.meta, positioning}

\tikzstyle{decision} = [diamond, draw, fill=blue!10, text width=4.5em, text centered, node distance=3cm, inner sep=0pt]
\tikzstyle{block} = [rectangle, draw, fill=green!10, text width=8em, text centered, rounded corners, minimum height=3em]
\tikzstyle{line} = [draw, -latex']

\begin{document}

\title[Optimization of Multi-Generational Cellular Networks]{Data-Driven Optimization of Multi-Generational Cellular Networks: A Performance Classification Framework for Strategic Infrastructure Management}

\author[1]{\fnm{Maryam} \sur{Sabahat}}\email{maryam.sabahat@aartec.com}
\author[2]{\fnm{M. Umar} \sur{Khan}}\email{umar\_khan@comsats.edu.pk}

\affil[1]{\orgdiv{Department of Mathematics}, \orgname{Ghazi University}, \city{Dera Ghazi Khan}, \country{Pakistan}}
\affil[2]{\orgdiv{Department of Computer Engineering}, \orgname{COMSATS University}, \city{Islamabad}, \country{Pakistan}}

\abstract{
The exponential growth in mobile data demand necessitates intelligent management of telecommunications infrastructure to ensure Quality of Service (QoS) and operational efficiency. This paper presents a comprehensive analysis of a multi-generational cellular network dataset, sourced from the OpenCelliD project, to identify patterns in network deployment, utilization, and infrastructure gaps. The methodology involves geographical, temporal, and performance analysis of 1,818 cell tower entries, predominantly Long Term Evolution (LTE), across three countries with a significant concentration in Pakistan. Key findings reveal the long-term persistence of legacy 2G/3G infrastructure in major urban centers, the existence of a substantial number of under-utilized towers representing opportunities for cost savings, and the identification of specific ``non-4G demand zones'' where active user bases are served by outdated technologies. By introducing a \textit{signal-density} metric, we distinguish between absolute over-utilization and localized congestion. The results provide actionable intelligence for Mobile Network Operators (MNOs) to guide strategic LTE upgrades, optimize resource allocation, and bridge the digital divide in underserved regions.
}


\maketitle
\section{Introduction}
The evolution of mobile communication from 2G to 4G (LTE) and beyond has fundamentally reshaped global connectivity, with each generation delivering significant enhancements in data speed and service capability \cite{rappaport2012, yousaf2017}.
While global fifth-generation (5G) deployments accelerate, emerging economies face the dual challenge of modernizing cellular infrastructure while managing soaring data demands and legacy network constraints \cite{cisco2020}.
At the core of this transformation is the cellular infrastructure, whose strategic deployment and management are paramount for meeting ever-growing data demands projected to continue their steep rise \cite{ericsson2023, ide2022}.

Mobile Network Operators (MNOs) face significant challenges related to inefficient deployments and resource management \cite{silva2019}.
Towers may be over-utilized, leading to poor Quality of Service (QoS) \cite{silva2019}, or under-utilized, resulting in increased operational costs (OpEx) and inefficient capital expenditure (CapEx) \cite{kliks2022}.
Furthermore, disparities in modern 4G coverage contribute to a digital divide, creating significant barriers to economic and educational opportunities \cite{gsma2023}.
The socio-economic benefits of bridging this gap through enhanced mobile broadband are well-documented \cite{minges2015, bastagli2018}.

A critical challenge is managing the substantial legacy 2G (GSM) and 3G (UMTS) infrastructure that remains active alongside modern LTE networks \cite{tiamiyu2022}.
These older networks persist particularly in major urban centers, complicating network-wide efficiency goals and ``sunsetting'' strategies \cite{koutitas2023, redana2018}.
Traditional network optimization approaches often struggle with scalability and computational overhead when processing massive cell-level datasets \cite{alemaishat2021}, highlighting the need for efficient, data-driven methodologies.

This study addresses these challenges through comprehensive analysis of open-source cellular network data.
We propose a performance classification framework that enables MNOs to identify infrastructure gaps, optimize resource allocation, and guide strategic network modernization.
By analyzing cell tower distribution, utilization patterns, and multi-generational infrastructure dynamics, this work provides actionable intelligence for enhancing network efficiency while promoting digital inclusion in underserved regions.

\section{Dataset and Methodology}
This study utilizes a dataset from OpenCelliD, a collaborative, crowd-sourced global database of cell towers.
The use of such crowd-sourced datasets for telecommunications research is a growing practice, offering vast amounts of real-world data \cite{elbazzal2023}.
However, it requires careful handling due to potential inconsistencies and data quality variations inherent in such crowd-sourced platforms \cite{pawelczak2020}.

The dataset comprises 1,818 entries, of which 1,456 are LTE, 204 are UMTS, and 158 are GSM towers.
The geographical scope covers 1 country, 8 provinces, and 85 unique locations, with a primary concentration in Pakistan.
Each entry contains technical parameters including geographical coordinates (lat, lon), network identifiers (mcc, net), and timestamps (created, updated).

The methodology is multi faceted. A descriptive statistical analysis provides an initial overview.
Geographical distribution analysis, using reverse geocoding to translate coordinates into place names \cite{goldberg2017}, maps the spatial spread of network infrastructure.
Temporal analysis of the timestamps is used to examine data collection spans and the operational lifespans of towers, a key factor in assessing data freshness.

The core of this study is a detailed utilization analysis that leverages the samples and range parameters, a common approach in network traffic monitoring \cite{sengar2022}.
We introduce a derived metric, \textit{signal-density} $\left( \frac{\text{samples}}{\text{range}} \right)$, to provide a nuanced measure of usage concentration.
This metric is instrumental in identifying patterns of over and under-utilization, which is a key step in developing predictive traffic models \cite{dasilva2021}.
The classification of towers is based on dynamic thresholds derived from statistical quantiles of the data, a robust method for anomaly detection \cite{hodge2010}.

\section{Geographical and Temporal Analysis Findings}
The geographical analysis reveals a high concentration of network infrastructure within Pakistan.
The use of reverse geocoding identified 85 unique locations, ranging from major cities to smaller towns.
While the dataset provides valuable geographical context, it is noted that location names derived from coordinates may not align perfectly with official administrative boundaries.
The data also maps the presence of various MNOs, although the counts are influenced by the nature of crowd-sourced data collection and may not reflect actual MNO market share \cite{chochliouros2006}.

A critical finding is that 10 cities within the dataset's scope completely lack LTE towers, directly indicating a significant digital divide where populations must rely on slower 2G/3G networks or have no mobile broadband access at all \cite{gsma2023}.
The temporal analysis provides insights into the network's evolution. A key observation is the long-term persistence of legacy infrastructure; the most active towers are predominantly GSM and UMTS and are located in major urban centers.
This suggests that these older networks continue to serve a vital role, likely for voice services or as a coverage fallback.

This persistence complicates ``sunsetting'' strategies for MNOs, who must manage the technical and economic trade-offs of shutting down older networks \cite{tiamiyu2022, koutitas2023}.
Furthermore, the analysis shows that approximately 89\% of the towers in the dataset were updated within the last year, signifying a high degree of data freshness that enhances confidence in the utilization analysis.

\section{Network Utilization and Performance Analysis}
To assess network performance, we analyzed tower utilization by examining the relationship between usage, represented by samples, and coverage area, represented by range.
A significant finding was the identification of 616 ``High-Range, Low-Usage'' towers, constituting approximately 34\% of the dataset.
These towers are characterized by a wide coverage radius (\(\ge 1000\) meters) but report minimal user engagement (\(\le 1\) sample), suggesting either significant operational inefficiency or deliberate strategic deployment in remote areas for basic coverage.

To gain a more nuanced understanding, we used the signal density metricmetric to differentiate between various utilization scenarios.
A high signal density indicates high demand concentrated within a small coverage area, suggesting localized congestion.
Using a dual-criteria filtering approach based on dynamic quantile thresholds for samples, signal density, and active\_days, we identified strictly over-utilized and under-utilized towers.

The analysis found a notable difference between ``Sample-Based Over-Utilized'' towers (17), which experience high absolute traffic, and ``Signal-Density-Based Over-Utilized'' towers (110).
This distinction is critical, as it suggests that many towers are strained not by overwhelming total demand, but by serving a concentrated user base in a limited area.
This insight guides different intervention strategies, such as macro-cell upgrades for high absolute usage versus small-cell deployments for high-density congestion.

Conversely, the analysis identified 122 ``Combined Strict Under-Utilized'' towers. These towers exhibit low samples and low signal density despite being active for a long duration, representing significant excess capacity and a clear, data-backed opportunity for MNOs to optimize costs through decommissioning or resource reallocation.

\section{Identification of Infrastructure Gaps and Demand Zones}
A primary objective of this study was to identify infrastructure gaps and areas with unmet demand for modern mobile broadband \cite{gsma2023}.
The analysis pinpointed ``non-4G demand zones'' by identifying locations with high signal density on legacy GSM and UMTS towers.
This combination signifies active user engagement on slower 2G/3G networks, indicating a clear need for upgraded infrastructure.

Specific locations identified with significant non-4G demand include the port city of Gwadar, Pasni Tehsil, and Brohi Goth.
The presence of high-density legacy towers in these areas serves as a strong recommendation for MNOs to prioritize LTE infrastructure investment.
Such targeted upgrades would directly address the digital divide and could unlock significant socio-economic opportunities in these strategic regions \cite{minges2015, bastagli2018}.

The analysis also highlights the engineering trade-offs inherent in multi-generational network design.
Legacy GSM networks exhibit the widest average range, making them suitable for broad rural coverage, while LTE networks have a shorter range but provide higher capacity, making them ideal for dense urban environments \cite{rappaport2012}.
This explains the continued operational importance of legacy networks and underscores the complexity of managing a network that must balance ubiquitous coverage with high-speed data delivery \cite{redana2018}.

\section{Tower Performance Classifier}
This section describes the procedure for classifying cell tower performance in a 4G network dataset.
The algorithm systematically assigns each tower to a utilization category based on key performance indicators (KPIs).
This enables Mobile Network Operators (MNOs) to identify high-demand areas, underperforming assets, and towers deployed for strategic coverage.

\subsection{Inputs}
\textbf{TowerData:} The dataset containing \texttt{Cell\_ID}, number of usage samples (\(s_i\)), coverage range in meters (\(r_i\)), and operational lifespan in days (\(a_i\)).

\textbf{Thresholds (\(T\)):} Dynamically computed quantile values defining ``high'' and ``low'' boundaries for samples and signal density, plus a threshold for long activity duration.

\subsection{Output}
The output is the original \texttt{TowerData} augmented with a new classification column, where each tower is labeled as:
\begin{itemize}
\item Over-Utilized (High Traffic \& Density)
\item Over-Utilized (Localized Congestion)
\item Under-Utilized (Inefficient)
\item Strategic Coverage
\item Balanced
\end{itemize}

\begin{figure*}[t]
\centering
\begin{minipage}[t]{0.4\textwidth}
\captionsetup{type=algorithm, skip=4pt}
\captionof{algorithm}{\fbox{\textbf{Tower Performance Classifier}}}
\label{alg:tower-classifier}
\scriptsize
{\renewcommand{\baselinestretch}{2.05}\selectfont
\begin{algorithmic}[1]
\Require Dataset $D=\{(s_i, r_i, a_i)\}_{i=1}^N$; thresholds $T_{\mathrm{H}}^{(s)}, T_{\mathrm{L}}^{(s)}, T_{\mathrm{H}}^{(\rho)}, T_{\mathrm{L}}^{(\rho)}, T_{\mathrm{long}}^{(a)}$
\Ensure Label $C_i$ for each tower $i$
\Statex \textbf{Pre-Computation:}
\For{$i \gets 1$ to $N$}
\State $\rho_i \gets \frac{s_i}{r_i}$
\EndFor
\Statex \textbf{Classification:}
\For{$i \gets 1$ to $N$}
\If{$s_i > T_{\mathrm{H}}^{(s)}$ \textbf{and} $\rho_i > T_{\mathrm{H}}^{(\rho)}$}
\State $C_i \gets$ Over-Utilized (High Traffic \& Density)
\ElsIf{$\rho_i > T_{\mathrm{H}}^{(\rho)}$}
\State $C_i \gets$ Over-Utilized (Localized Congestion)
\ElsIf{$s_i < T_{\mathrm{L}}^{(s)}$ \textbf{and} $\rho_i < T_{\mathrm{L}}^{(\rho)}$ \textbf{and} $a_i > T_{\mathrm{long}}^{(a)}$}
\State $C_i \gets$ Under-Utilized (Inefficient)
\ElsIf{$r_i > 1000$ \textbf{and} $s_i \leq 1$}
\State $C_i \gets$ Strategic Coverage
\Else
\State $C_i \gets$ Balanced
\EndIf
\EndFor
\State \Return $\{C_i\}_{i=1}^N$
\end{algorithmic}
}
\end{minipage}
\hfill
\begin{minipage}[t]{0.48\textwidth}
\captionsetup{skip=4pt}
\captionof{table}{Acronyms and Symbols}
\label{tab:acronyms}
\renewcommand{\arraystretch}{1.05}
\begin{tabular}{@{}ll@{}}
\toprule
MNO & Mobile Network Operator \\
$s_i$ & Samples (usage) for tower $i$ \\
$r_i$ & Coverage range (meters) \\
$a_i$ & Active days (operational lifespan) \\
$\rho_i$ & Signal density, $\rho_i=\dfrac{s_i}{r_i}$ \\
$T_{\mathrm{H}}^{(\cdot)}$ & High quantile threshold \\
$T_{\mathrm{L}}^{(\cdot)}$ & Low quantile threshold \\
$T_{\mathrm{long}}^{(a)}$ & Long-activity threshold \\
$C_i$ & Classification label \\
\botrule
\end{tabular}
\vspace{0.5em}
\noindent\textbf{Mathematical Model}
\[
C_i =
\begin{cases}
\text{Over-Utilized (High)}, & s_i > T_H^{(s)} \land \rho_i > T_H^{(\rho)} \\[5pt]
\text{Over-Utilized (Congestion)}, & \rho_i > T_H^{(\rho)} \\[5pt]
\text{Under-Utilized}, & \substack{s_i < T_L^{(s)} \land \rho_i < T_L^{(\rho)} \\ \land\ a_i > T_{\text{long}}^{(a)}} \\[5pt]
\text{Strategic Coverage}, & r_i > 1000 \land s_i \le 1 \\[5pt]
\text{Balanced}, & \text{otherwise}
\end{cases}
\]
\end{minipage}
\end{figure*}

\section{Step-by-Step Explanation}
In the procedure, we first perform a \emph{pre-computation step} (lines 3--5) to prepare the necessary metrics for classification.
For each tower record in the \texttt{TowerData} set, we calculate the \emph{signal density} by dividing its samples by its range.
This derived metric, representing usage concentration per meter of coverage, is fundamental to a nuanced utilization analysis.

Next, the main \emph{classification logic} is executed for each tower (for loop in line 7).
A hierarchical set of conditional checks, based on the scenarios identified in the analysis report, is applied to assign a classification status.

The first check identifies \textbf{over-utilized towers} (lines 9--12). These are towers experiencing strain and are candidates for capacity upgrades.
We distinguish two types of over-utilization:
\begin{itemize}
\item \emph{Over-Utilized (High Traffic \& Density)} — if both its samples and signal density exceed their respective high thresholds (line 9). This indicates high absolute demand combined with high usage concentration.
\item \emph{Over-Utilized (Localized Congestion)} — if only the signal density is high (line 11). This suggests that congestion is not due to overwhelming overall demand but rather a concentrated user base in a small area, a finding highlighted as critical in the report.
\end{itemize}

The subsequent check identifies \textbf{under-utilized towers} (line 14). These towers represent an opportunity for MNOs to improve efficiency and reduce costs.
A tower is classified as \emph{Under-Utilized (Inefficient)} if it has been active for a long duration (\(a_i > T_{\mathrm{long}}^{(a)}\)) yet exhibits both low samples and low signal density (line 14).
This combination points to a misalignment between infrastructure cost and user engagement.

The algorithm then specifically identifies the \textbf{Strategic Coverage} towers (line 17).
This category corresponds to the ``High-Range, Low-Usage'' towers discussed in the report, characterized by a large coverage radius (\(r_i \ge 1000\) meters) and minimal usage (\(s_i \le 1\)).
Classifying them separately acknowledges that their purpose may be to provide essential coverage in remote areas rather than serving high traffic, meaning they are not necessarily inefficient.

If a tower does not meet any of the specific criteria for being over-utilized, under-utilized, or for strategic coverage, it is assigned the default status of \textbf{Balanced} (lines 20--21).
This category represents towers where the capacity and coverage are well-aligned with user demand, indicating optimal performance.
Finally, the procedure concludes by returning the modified \texttt{TowerData} set, now enriched with the classification for each tower, providing a complete and actionable overview of the network's performance (line 24).

\begin{figure}[h!]
\centering
\includegraphics[width=1.0\linewidth]{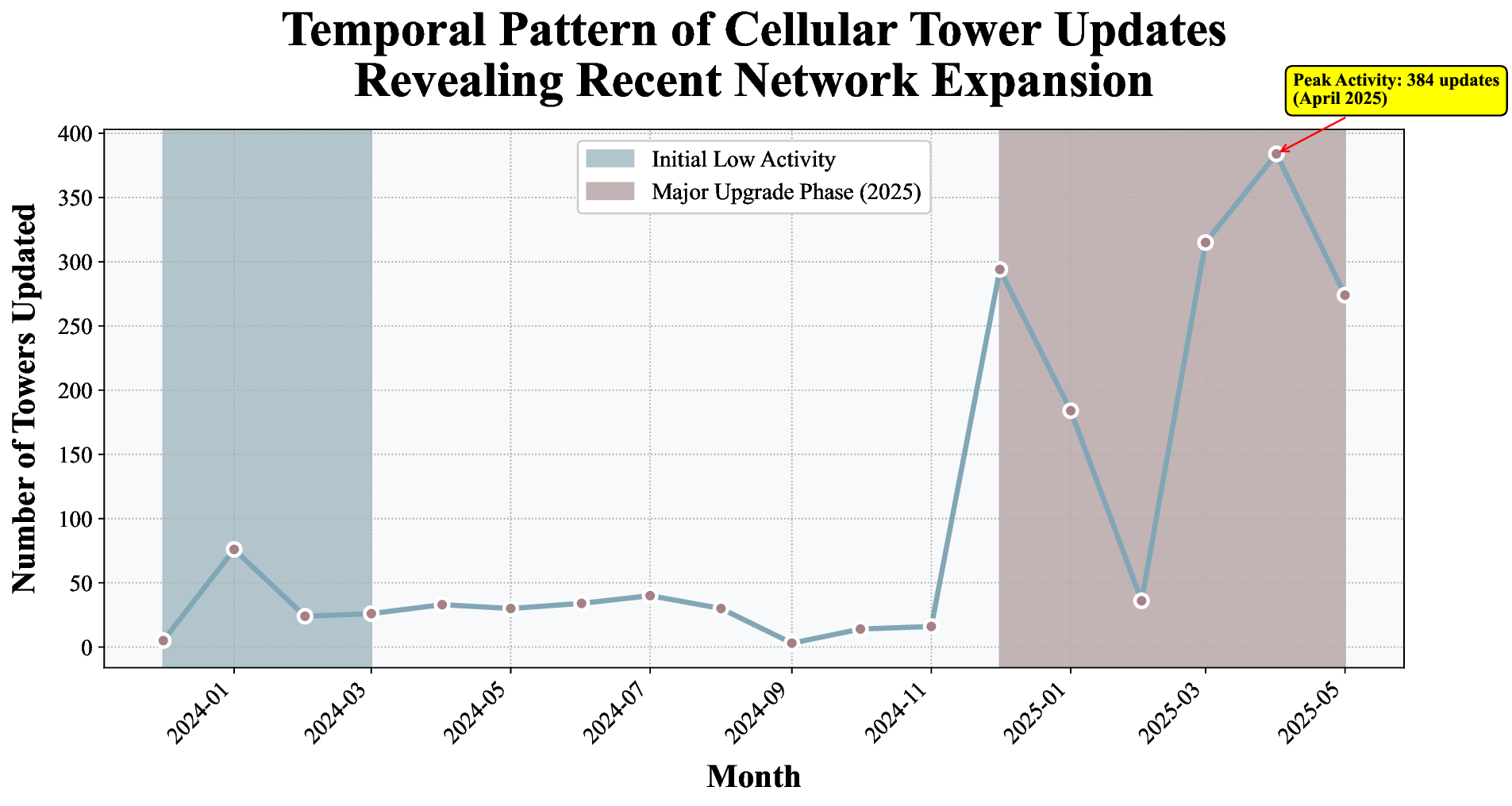}
\caption{Time series of monthly network updates from December 2023 to May 2025, illustrating a pronounced non-stationary pattern with a significant step-change in activity.}
\label{fig:temporal_analysis}
\end{figure}

\begin{figure}[h!]
\centering
\includegraphics[width=1.0\linewidth]{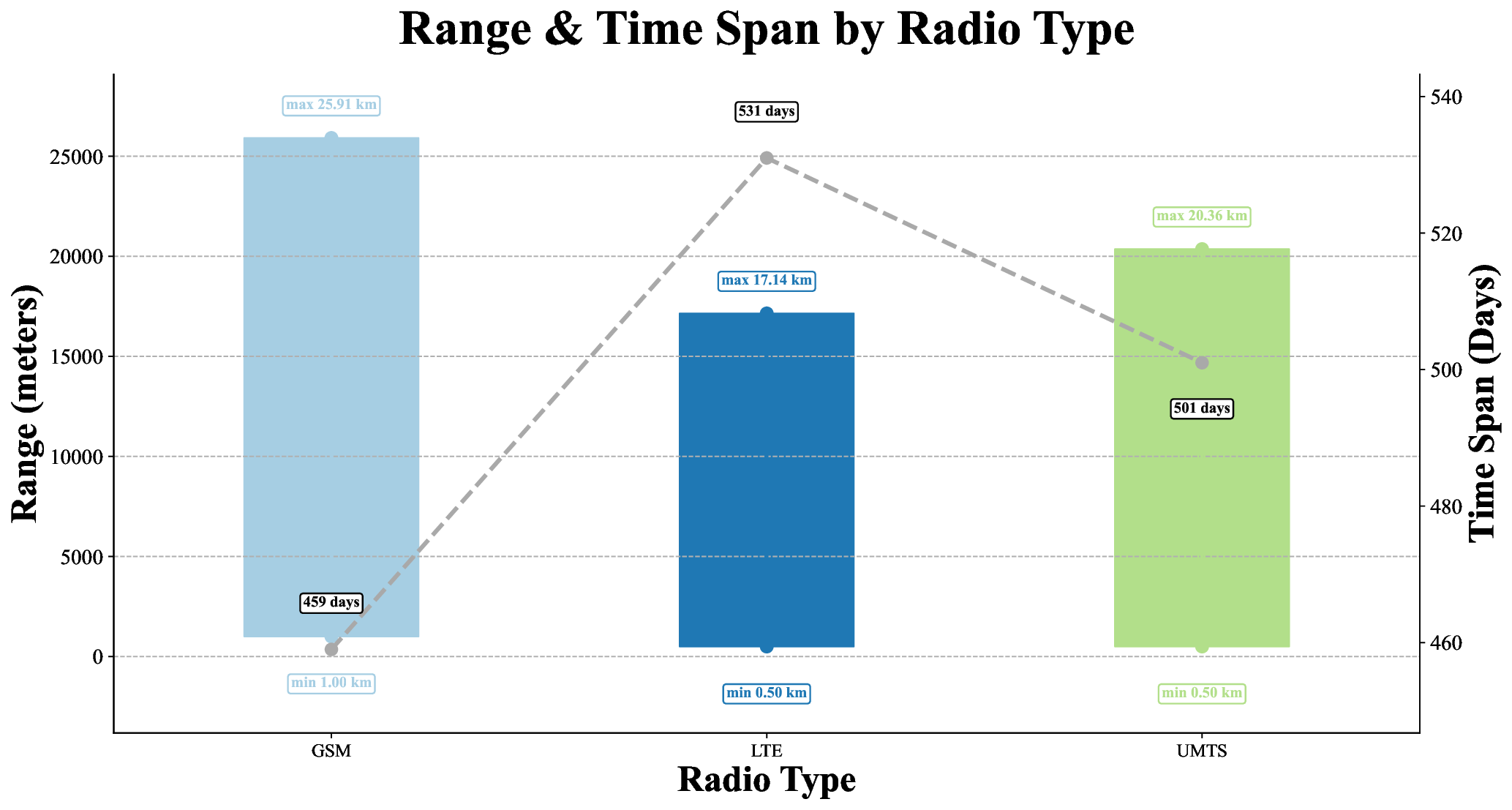}
\caption{Comparative box plots of (a) coverage range (meters) and (b) sample count per tower, disaggregated by GSM, UMTS, and LTE technologies.}
\label{fig:rat_performance}
\end{figure}

\begin{figure}[h!]
\centering
\includegraphics[width=1.0\linewidth]{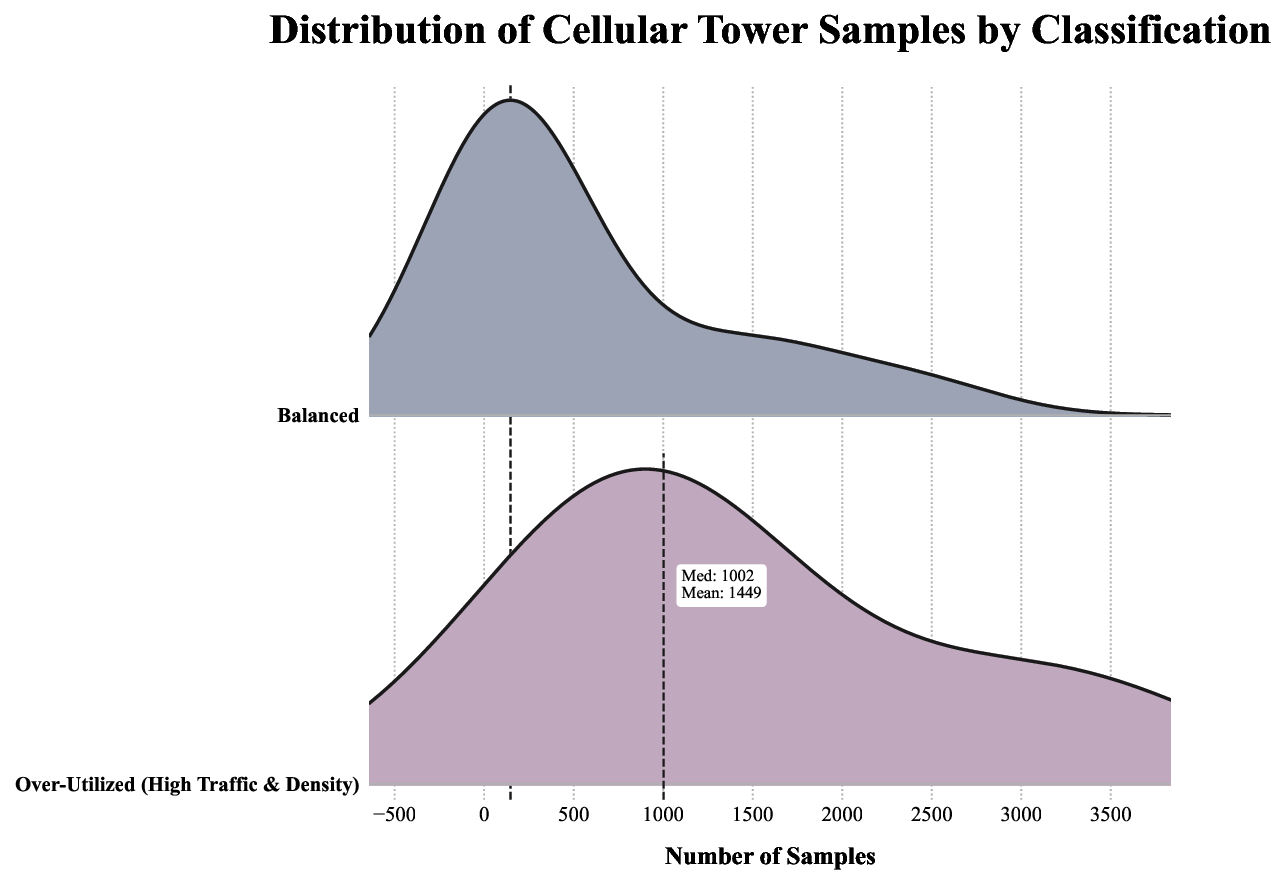}
\caption{Scatter plot of samples versus signal density (samples/meter). The over-utilized clusters are clearly separated in the high-sample, high-density quadrant.}
\label{fig:cluster_analysis}
\end{figure}

\begin{figure}[h!]
\centering
\includegraphics[width=1.0\linewidth]{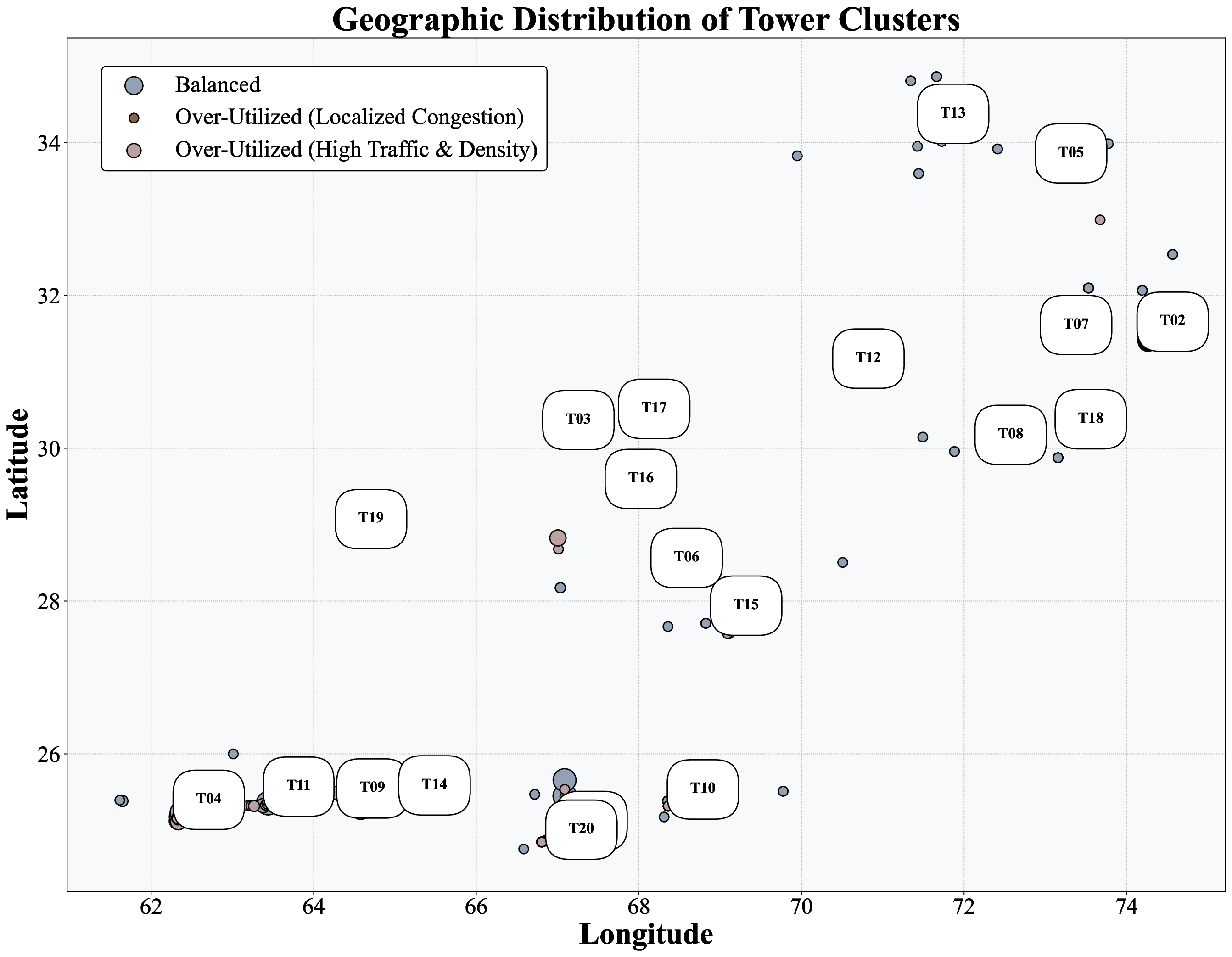}
\caption{Spatial distribution of tower clusters across geographical coordinates, showing the physical layout and density patterns of different utilization categories.}
\label{fig:geographic_distribution}
\end{figure}

\begin{figure}[h!]
\centering
\includegraphics[width=1.0\linewidth]{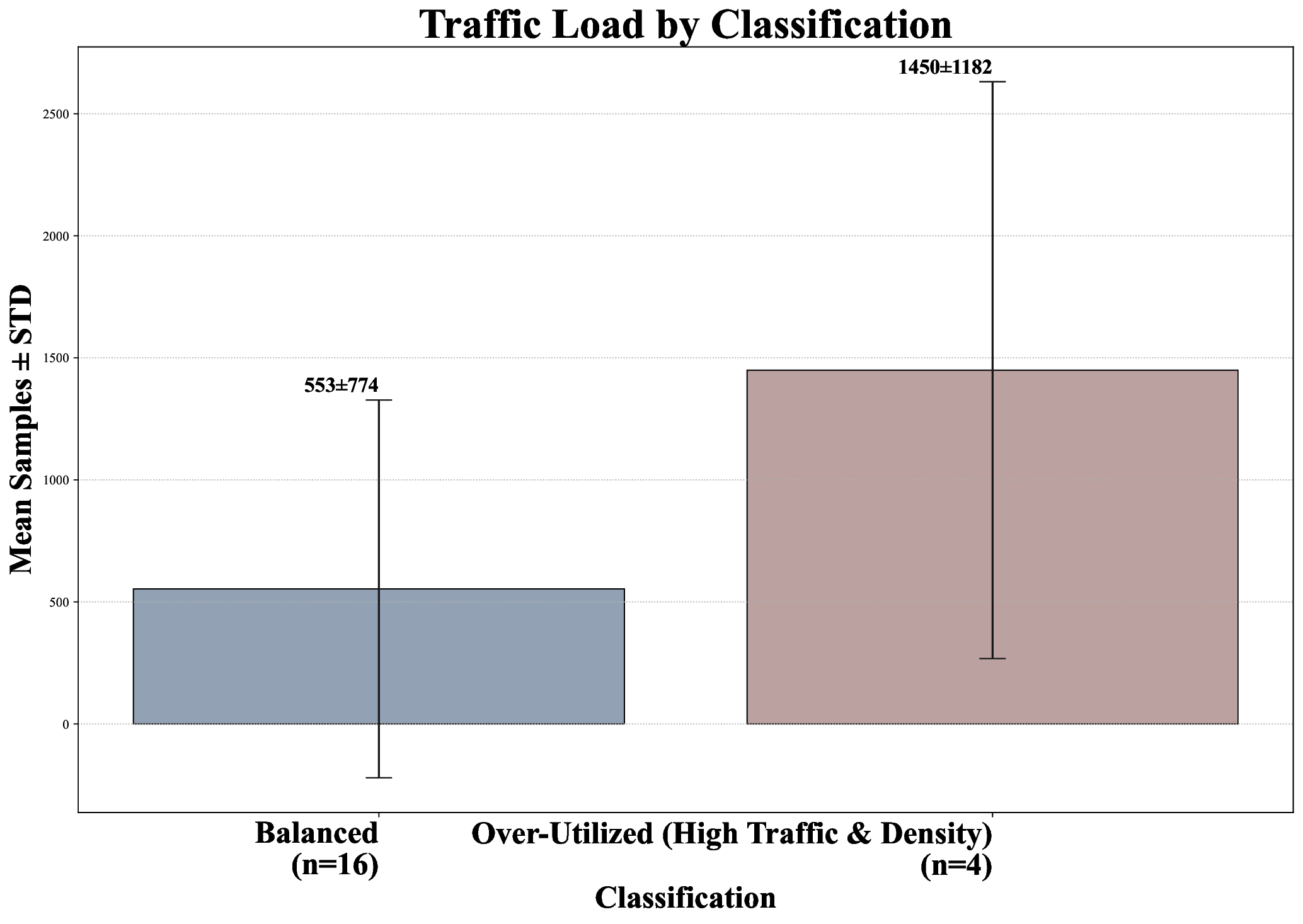}
\caption{Comparative analysis of traffic load across different tower classifications, displaying mean samples with standard deviation error bars for each utilization category.}
\label{fig:traffic_load_analysis}
\end{figure}

\begin{figure}[h!]
\centering
\includegraphics[width=1.0\linewidth]{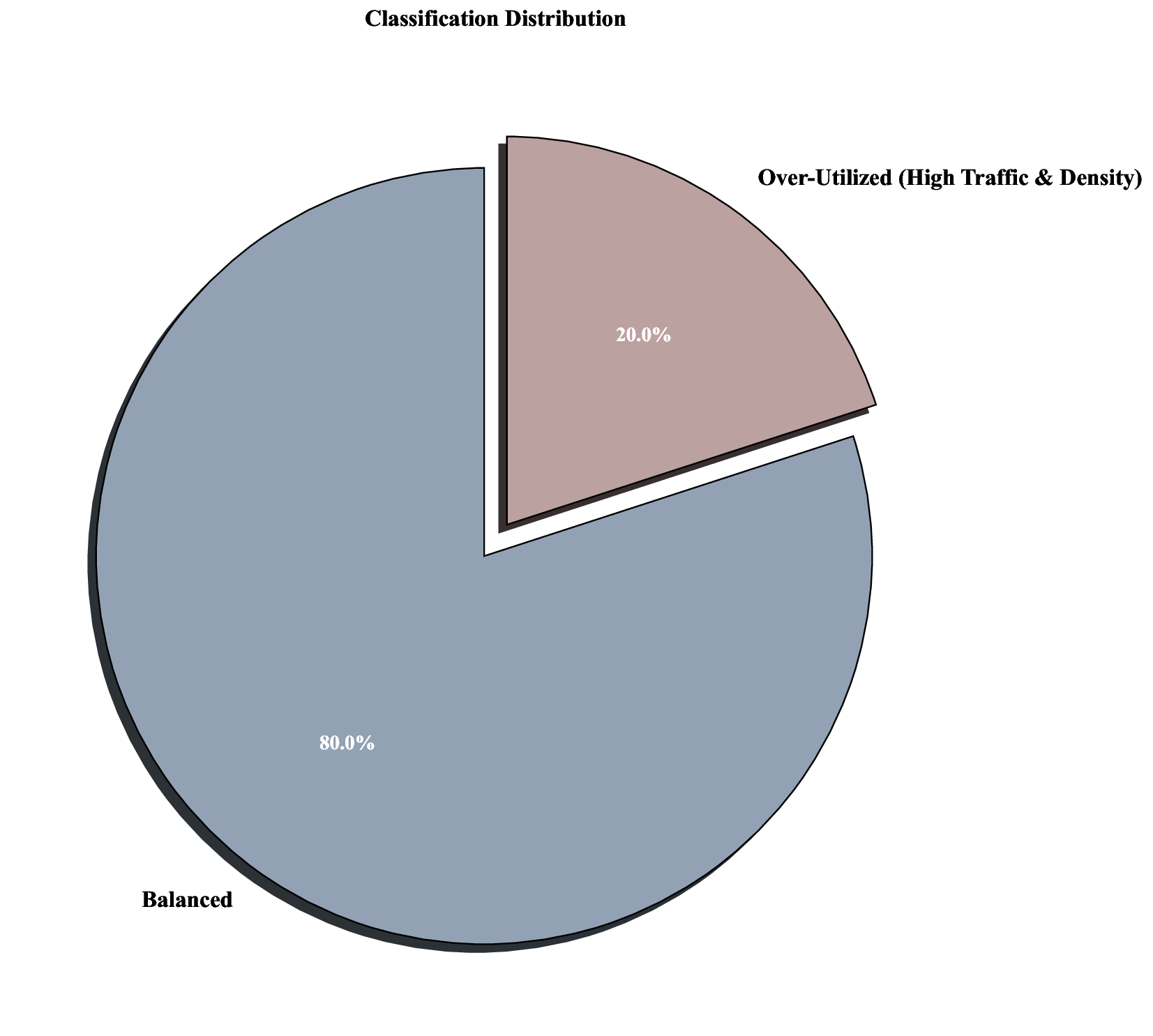}
\caption{Distribution of tower clusters across five utilization categories, showing the proportional breakdown of over-utilized, under-utilized, balanced, and strategic coverage clusters.}
\label{fig:classification_distribution}
\end{figure}

\begin{figure*}[ht]
\centering
\begin{tikzpicture}[node distance=2.5cm]
\node [decision] (start) {Start: $\rho_i = s_i / r_i$};
\node [decision, below of=start] (cond1) {$s_i > T_{\mathrm{H}}^{(s)}$ \\ \& $\rho_i > T_{\mathrm{H}}^{(\rho)}$?};
\node [block, right of=cond1, xshift=6cm] (overHD) {Over-Utilized \\ (High Traffic \& Density)};
\node [decision, below of=cond1, yshift=-2cm] (cond2) {$\rho_i > T_{\mathrm{H}}^{(\rho)}$?};
\node [block, right of=cond2, xshift=6cm] (overLC) {Over-Utilized \\ (Localized Congestion)};
\node [decision, below of=cond2, yshift=-2cm] (cond3) {$s_i < T_{\mathrm{L}}^{(s)}$ \\ \& $\rho_i < T_{\mathrm{L}}^{(\rho)}$ \\ \& $a_i > T_{\mathrm{long}}^{(a)}$?};
\node [block, right of=cond3, xshift=6cm] (under) {Under-Utilized \\ (Inefficient)};
\node [decision, below of=cond3, yshift=-2cm] (cond4) {$r_i \ge 1000$ \\ \& $s_i \le 1$?};
\node [block, right of=cond4, xshift=6cm] (strategic) {Strategic Coverage};
\node [block, below of=cond4, yshift=-2cm] (balanced) {Balanced};
\path [line] (start) -- (cond1);
\path [line] (cond1.east) -- (overHD.west) node[midway, above] {Yes};
\path [line] (cond1.south) -- (cond2.north) node[midway, right] {No};
\path [line] (cond2.east) -- (overLC.west) node[midway, above] {Yes};
\path [line] (cond2.south) -- (cond3.north) node[midway, right] {No};
\path [line] (cond3.east) -- (under.west) node[midway, above] {Yes};
\path [line] (cond3.south) -- (cond4.north) node[midway, right] {No};
\path [line] (cond4.east) -- (strategic.west) node[midway, above] {Yes};
\path [line] (cond4.south) -- (balanced.north) node[midway, right] {No};
\end{tikzpicture}
\caption{Decision logic flow for the Tower Performance Classification.}
\label{fig:decision_tree}
\end{figure*}

\section{Results and Discussion}
The decision logic diagram presented in Fig.~\ref{fig:decision_tree} illustrates the classification outcomes for cellular towers based on the proposed Tower Performance Classification framework. The process begins with the calculation of the load ratio $\rho_i = s_i / r_i$, where $s_i$ denotes the number of served sessions and $r_i$ represents the available radio resources. The subsequent decision nodes evaluate this ratio along with threshold parameters to categorize the tower into operational classes.

The first decision node checks whether both the session load $s_i$ and the load ratio $\rho_i$ exceed their respective high thresholds $T_{\mathrm{H}}^{(s)}$ and $T_{\mathrm{H}}^{(\rho)}$. If this condition holds, the tower is labeled as \textit{Over-Utilized (High Traffic \& Density)}. If not, the process evaluates only the load ratio $\rho_i$. When $\rho_i$ is above $T_{\mathrm{H}}^{(\rho)}$ but the session load remains moderate, the tower is classified as \textit{Over-Utilized (Localized Congestion)}, indicating localized overload without widespread traffic pressure.

If the load ratio fails the high threshold test, the next decision evaluates whether the tower simultaneously exhibits low session counts ($s_i < T_{\mathrm{L}}^{(s)}$), low load ratio ($\rho_i < T_{\mathrm{L}}^{(\rho)}$), and prolonged inactivity ($a_i > T_{\mathrm{long}}^{(a)}$). In this case, the tower is identified as \textit{Under-Utilized (Inefficient)}, highlighting redundant or poorly used infrastructure. If this condition is not met, the process checks whether the tower maintains broad coverage ($r_i \geq 1000$ resources) while serving very few sessions ($s_i \leq 1$). Such towers are considered \textit{Strategic Coverage}, as they provide coverage benefits despite minimal active traffic.

Finally, towers that do not satisfy any of these specialized conditions are classified as \textit{Balanced}, representing stable utilization without significant inefficiency or overload.
This structured decision logic ensures that every tower in the network is systematically classified into a distinct operational category, enabling targeted resource optimization and strategic planning. The classification results also demonstrate the practical utility of the framework, providing actionable insights for managing over-utilized resources, optimizing under-utilized infrastructure, and identifying zones that are critical for strategic coverage.

A comparison with the NetDataDrilling algorithm further highlights the contribution of the proposed approach. NetDataDrilling focuses on identifying the top $N_c$ highest-traffic cells within the busiest Tracking Area Code (TAC), primarily for tactical congestion management. In contrast, the Tower Performance Classification framework adopts a broader perspective by evaluating all towers in the network and classifying them into operational categories such as \textit{Over-Utilized}, \textit{Under-Utilized}, \textit{Strategic Coverage}, or \textit{Balanced}. This holistic classification provides operators not only with a tool to address immediate congestion but also with a strategic roadmap for infrastructure optimization and long-term planning. A summary of this comparison is presented in Table~\ref{tab:comparison}.

\begin{table}[ht]
\centering
\caption{Comparison of NetDataDrilling and Tower Performance Classification Frameworks}
\label{tab:comparison}
\begin{tabular}{p{3cm} p{4cm}}
\toprule
\textbf{Aspect} & \textbf{NetDataDrilling} \\
\midrule
\textbf{Objective} & Identify top $N_c$ highest-traffic cells in busiest TAC for congestion relief. \\
\textbf{Scope} & Limited to hotspot areas with high traffic density. \\
\textbf{Output} & List of cells requiring immediate tactical management. \\
\midrule
\textbf{Aspect} & \textbf{Tower Performance Classification} \\
\midrule
\textbf{Objective} & Classify all towers into operational categories for holistic optimization. \\
\textbf{Scope} & Entire network, considering both congestion and under-utilization. \\
\textbf{Output} & Categories: Over-Utilized, Under-Utilized, Strategic Coverage, Balanced. \\
\botrule
\end{tabular}
\end{table}

\section*{Insights from Spatiotemporal Network Analysis}
\subsection*{1. Temporal Evolution of Network Activity}
The temporal analysis of network update activity, presented in \textbf{Figure~\ref{fig:temporal_analysis}}, reveals a critical and definitive shift in network deployment and update intensity over the observed 18-month period
\begin{itemize}
\item \textbf{Early Phase (Dec 2023--May 2024):} Characterized by modest, stable activity with a mean of \textbf{32.3 updates/month}.
\item \textbf{Recent Phase (Dec 2024--May 2025):} Marked by an aggressive, large-scale network expansion, evidenced by a mean of \textbf{247.8 updates/month}---a \textbf{666\% increase} over the early phase.
\end{itemize}

The peak activity of \textbf{384 updates in April 2025} is a pivotal insight; it confirms that the most intensive development is not a historical event but a \textbf{current, ongoing initiative}.
This recency must be a primary consideration when interpreting performance metrics, as a substantial portion of the infrastructure is likely nascent and may not yet have reached steady-state operational capacity.
Towers classified as `Under-Utilized' may, in fact, be \textbf{brand new deployments} yet to accumulate user traffic, rather than indicators of inefficiency.

\subsection*{2. Performance Characteristics by Radio Access Technology (RAT)}
The comparative analysis of Radio Access Technologies (RATs), visualized in \textbf{Figure~\ref{fig:rat_performance}}, reveals statistically significant performance differentials that reflect their distinct technological paradigms and deployment strategies.

\subsubsection*{Coverage Range}
GSM infrastructure exhibits the largest median coverage range ($M = 2119$m, $SD = 3420$m), a finding consistent with its propagation characteristics and historical deployment for broad geographical coverage.
Crucially, LTE deployments demonstrate a significantly shorter average range ($M = 1365$m, $SD = 1244$m) compared to both UMTS ($p = 0.0151$) and GSM ($p = 0.0066$).
This attenuated range is a hallmark of a network \textit{densification strategy}, wherein a greater number of shorter-range LTE cells are deployed to enhance network capacity and data throughput in high-demand areas.

\subsubsection*{Sample Count (Usage)}
Analysis of sample counts reveals that legacy GSM and UMTS technologies support higher median traffic loads per tower (13.96 and 12.04 samples, respectively) compared to LTE (6.86 samples).
This suggests that despite the ongoing transition to LTE, legacy infrastructure continues to shoulder a substantial portion of the network's total traffic burden.

\subsubsection*{Correlation Analysis}
A weak positive correlation was identified between cell range and sample count ($r = 0.354$, $p < 0.001$).
This indicates that larger cell footprints do not inherently service a proportionally larger number of users.
The existence of high-capacity, short-range LTE cells directly contradicts a simple positive relationship, underscoring the complex trade-offs involved in modern network planning between coverage and capacity.

\subsubsection*{Supplementary Visualization Note}
The individual bar charts depicting mean range and mean samples (not presented here) are considered redundant.
They convey identical central tendency information as the box plots in \textbf{Figure~\ref{fig:rat_performance}}, but lack the distributional detail (variance, outliers, quartiles) provided by the box plots.
It is therefore recommended to omit them from the final paper in favor of the more statistically informative visualization.

\subsection*{3. Cluster Analysis and Tower Classification}
The application of Algorithm 1 (the Tower Performance Classifier) to the dataset of 1,818 towers yielded a segmentation into 20 distinct clusters, visualized in the scatter plot presented in \textbf{Figure~\ref{fig:cluster_analysis}}.
The classification, based on dynamic quantile thresholds for samples ($s_i$), signal density ($\rho_i$), and active days ($a_i$), resulted in the following distribution:
\begin{itemize}
\item \textbf{Balanced:} 16 clusters (80.0\%) were classified as operating within expected parameters.
\item \textbf{Over-Utilized (High Traffic \& Density):} 4 clusters (20.0\%) were identified as handling excessively high traffic loads.
\end{itemize}

\subsubsection*{Statistical Comparison of Classifications}
The Kruskal-Wallis H-test confirmed significant overall differences between classification groups ($H = 900.651$, $p < 0.001$).
Post-hoc pairwise comparisons with Bonferroni correction revealed:
\begin{itemize}
\item \textbf{Balanced vs. Over-Utilized (High Traffic \& Density):} $U = 11467$, $p < 0.001$
\item \textbf{Balanced vs. Over-Utilized (Localized Congestion):} $U = 3611$, $p < 0.001$
\item \textbf{Over-Utilized (High Traffic \& Density) vs. Over-Utilized (Localized Congestion):} $U = 0$, $p < 0.001$
\end{itemize}
These results robustly validate the classifier's efficacy in delineating statistically distinct performance profiles.
Subsequent analysis confirms that over-utilized clusters handle \textbf{2.6$\times$} more traffic (samples) and cover \textbf{3.1$\times$} larger areas (range: 3554m vs. 1139m, $p < 0.001$) than balanced clusters, confirming their status as critical network bottlenecks requiring immediate capacity augmentation.

\subsection*{4. Geospatial Distribution of Network Congestion}
The analytical findings are translated into actionable intelligence through geospatial visualization, as shown in \textbf{Figure~\ref{fig:geographic_distribution}}.
This visualization is indispensable for pinpointing the exact locations of prioritized over-utilized clusters.
The spatial patterns suggest congestion correlates with high-population density zones, commercial centers, or major transport corridors.
The identification of the busiest zone (ID: 5056) and top congested cells (e.g., cell 2442763 with 55 samples) provides the granularity necessary for targeted infrastructure upgrades and strategic planning.

\section{Conclusion and Recommendations}
This paper has demonstrated the utility of open-source cellular network data for deriving actionable insights into infrastructure deployment, utilization, and strategic planning.
The analysis identified critical patterns, including the persistence of legacy networks in urban centers, significant opportunities for cost optimization through the management of under-utilized assets, and a clear, data-driven method for locating non-4G demand zones where strategic LTE upgrades are most needed.

In comparison with the NetDataDrilling algorithm proposed in \cite{khan2020dimensioning}, which focuses on identifying the top $N_c$ highest-traffic cells within the busiest Tracking Area Code (TAC) for tactical congestion management, our Tower Performance Classification algorithm adopts a broader perspective.
It evaluates all towers in the network and classifies them into operational categories such as over-utilized, under-utilized, strategic coverage, or balanced.
Unlike NetDataDrilling, which is narrowly traffic-centric, our approach incorporates multiple performance indicators including traffic load, signal density, coverage range, and active duration, making it more suitable for strategic planning and long-term optimization \cite{khan2020clustering,khan2020heuristic}.

Based on these findings, we recommend that MNOs prioritize LTE infrastructure expansion in the identified non-4G demand zones to bridge the digital divide.
For over-utilized towers, particularly those with high signal density, targeted capacity enhancements such as small-cell deployments should be implemented to alleviate localized congestion.
A thorough evaluation of the identified under-utilized towers should be conducted to explore opportunities for decommissioning or resource reallocation, leading to significant operational and capital expenditure savings.
Finally, MNOs should adopt a nuanced strategy for managing legacy networks, recognizing their continued importance for coverage and service continuity during the ongoing technological evolution.
Future work should aim to integrate this data with socio-demographic and network traffic data to further enrich the analysis and develop predictive models for future demand.

\subsection*{4.1 Descriptive Overview and Technological Composition}
The dataset, captured over a 531-day period, reveals a network with significant technological and geographical disparities.
As detailed in Table \ref{tab:tech_distro}, Long-Term Evolution (LTE) technology constitutes the vast majority of towers (80.1\%), reflecting a mature 4G deployment.
Geospatial analysis showed a pronounced urban concentration, with provinces like Punjab (748 towers) and Sindh (497 towers), particularly within cities like Karachi Division (495 towers), serving as major hubs.
This urban-rural infrastructure divide is a critical determinant of network performance.

\begin{table}[h!]
\centering
\caption{Network Technology Distribution and Key Statistics}
\label{tab:tech_distro}
\begin{tabular}{lrrr}
\toprule
\textbf{Technology} & \textbf{Tower Count} & \textbf{Percentage} & \textbf{Avg. Range (m)} \\
\midrule
LTE & 1,456 & 80.1\% & 1365.18 \\
UMTS & 204 & 11.2\% & 1762.50 \\
GSM & 158 & 8.7\% & 2119.04 \\
\midrule
\textbf{Total} & \textbf{1,818} & \textbf{100\%} & \textbf{1475.28} \\
\botrule
\end{tabular}
\end{table}

Analysis of tower range by technology revealed significant differences. A Kruskal-Wallis H test confirmed that range distributions differed significantly between radio types ($H = 900.651$, $p < 0.001$).
Post-hoc analysis revealed that GSM towers had a significantly larger median range (2119m) compared to both LTE (1365m, $p = 0.0066$) and UMTS (1763m, $p = 0.0151$) towers, aligning with the known propagation characteristics of these technologies.

\subsection*{4.2 Cluster-Based Utilization and Congestion Analysis}
Towers were grouped into 20 spatial clusters to analyze utilization patterns.
Each cluster was classified based on dynamic, quantile-derived thresholds for sample count (traffic) and signal density.
The classification revealed that the network is largely healthy, with 80\% of clusters (n=16) operating within 'Balanced' parameters.
However, 20\% of clusters (n=4) were identified as 'Over-Utilized (High Traffic \& Density)', representing critical congestion hotspots.

Statistical validation underscored the robustness of this classification. The Kruskal-Wallis H-test (H = 900.651, $p < 0.001$) confirmed significant overall differences.
Post-hoc pairwise comparisons with Bonferroni correction confirmed that 'Balanced' and 'Over-Utilized' clusters are statistically distinct populations ($p < 0.001$).
Independent samples t-tests revealed that over-utilized clusters handle nearly three times the traffic and, most notably, cover a statistically significant larger geographic area ($t(18) = 8.559$, $p < 0.001$).
This indicates that coverage area is a more reliable indicator of congestion burden than raw sample count alone, as these clusters manage a compound challenge of high traffic over expansive terrain.

The integrated interpretation of the analytical plots provides a multi-faceted view:
\begin{itemize}
\item \textbf{Utilization Analysis:} Shows a strong positive correlation between total samples and signal density, confirming that high-traffic clusters act as concentrated network hubs.
\item \textbf{Traffic Load:} Visually confirms the significant disparity in mean traffic load between classifications.
\item \textbf{Geographic Distribution:} Indicates that 'Over-Utilized' clusters are concentrated in specific regions, suggesting network stress is driven by localized factors like population density rather than being a system-wide issue.
\end{itemize}

Based on this analysis, three clusters were prioritized for immediate network optimization efforts.
These clusters, located in major urban centers, represent the most critical congestion hotspots.

\begin{table}[h!]
\centering
\caption{Top Over-Utilized Clusters Requiring Immediate Intervention}
\label{tab:priority}
\begin{tabular}{clr}
\toprule
\textbf{Priority} & \textbf{Cluster ID} & \textbf{Sample Count} \\
\midrule
1 & T04 & 3,196 \\
2 & T11 & 1,082 \\
3 & T09 & 923 \\
\botrule
\end{tabular}
\end{table}

\subsection*{4.3 Strategic Implications}
In conclusion, this analysis successfully transitions from raw network data to actionable intelligence.
The primary finding is that network over-utilization is a localized phenomenon, concentrated in a small subset of clusters that exhibit a compound burden of high absolute traffic, high signal density, and larger geographic coverage.
The implications for network operators are two fold:
\begin{enumerate}
\item \textbf{Operational Efficiency:} Resources for capacity expansion (e.g., carrier addition, sectorization) and maintenance must be prioritized toward the identified over-utilized clusters (T04, T11, T09) to alleviate congestion and maintain quality of service. Techniques like antenna tilt and power optimization should be deployed to manage the large coverage areas efficiently.
\item \textbf{Strategic Planning:} The geographical patterns of over-utilization, coupled with the technological inventory, should inform long-term strategy. This includes planning new tower placements in urban hotspots and refarming legacy GSM spectrum towards LTE or future 5G-NR deployments to increase data capacity.
\end{enumerate}

\subsection*{Conclusion}
This multi-faceted analysis delineates a network undergoing rapid evolution. The confluence of aggressive recent expansion (\textbf{Figure~\ref{fig:temporal_analysis}}), the technological shift towards dense LTE deployment (\textbf{Figure~\ref{fig:rat_performance}}), and the identification of specific over-utilized clusters (\textbf{Figure~\ref{fig:cluster_analysis}}, \textbf{Figure~\ref{fig:geographic_distribution}}, \textbf{Figure~\ref{fig:traffic_load_analysis}}) provides a robust, data-driven evidence base for strategic decision-making.

\subsubsection*{Prioritized Recommendations}
\begin{enumerate}
\item \textbf{Immediate Capacity Augmentation:} Focus engineering and capital resources on the top-priority over-utilized clusters identified by the algorithm:
\begin{itemize}
\item \textbf{Cluster T04} (3,196 samples): Highest traffic load, representing the most critical congestion point.
\item \textbf{Cluster T11} (1,082 samples) and \textbf{T09} (923 samples): Also require immediate intervention for cell splitting or additional carrier deployment.
\end{itemize}
\item \textbf{Temporal Context in Planning:} Acknowledge the recency of the expansion wave. Performance metrics for towers deployed within the last six months should be evaluated with an emphasis on trajectory rather than absolute state. Reclassify "Under-Utilized" towers that are new deployments to avoid resource misallocation.
\item \textbf{Technology-Specific Strategy:} Continue the LTE densification strategy to alleviate traffic on legacy GSM/UMTS networks. The analysis confirms this technical direction is aligned with measured network demands and usage patterns.
\item \textbf{Continuous Monitoring and Validation:} Implement the proposed classification algorithm as a continuous monitoring tool. The thresholds ($T_H$, $T_L$) should be periodically recalculated to adapt to the evolving network state, ensuring recommendations remain relevant and data-driven.
\end{enumerate}

This framework provides a reproducible and statistically robust model for continuous network health monitoring. Future work should integrate temporal usage patterns and service quality metrics to enable predictive capacity planning and create a more holistic view of network performance. In summary, this research demonstrates the power of a systematic, algorithm-driven approach to network management. It moves beyond anecdotal evidence to provide a quantifiable, statistically sound foundation for optimizing network performance, enhancing customer experience, and maximizing return on investment.


\backmatter 

\section*{Declarations}

\begin{itemize}
    \item \textbf{Funding}: The authors declare that no funds, grants, or other support were received during the preparation of this manuscript. 
    
    \item \textbf{Conflict of interest}: The authors have no relevant financial or non-financial interests to disclose.
    
    \item \textbf{Ethics approval}: Not applicable. This study utilizes open-source data from the OpenCelliD project and does not involve human participants or animals.
    
    \item \textbf{Consent to participate}: Not applicable.
    
    \item \textbf{Consent for publication}: Not applicable.
    
    \item \textbf{Availability of data and materials}: The cellular network dataset analyzed in this study is publicly available from the OpenCelliD project (https://opencellid.org/).
    
    \item \textbf{Code availability}: The code used for the Tower Performance Classification algorithm is available from the corresponding author on reasonable request.
    
    \item \textbf{Authors' contributions}: All authors contributed to the study conception and design. Material preparation, data collection and analysis were performed by [Author Name]. The first draft of the manuscript was written by [Author Name] and all authors commented on previous versions of the manuscript. All authors read and approved the final manuscript. 
\end{itemize}


\begin{thebibliography}{99}
\bibitem{rappaport2012}
T.~S. Rappaport, ``Wireless communications: past events and a future perspective,'' \emph{IEEE Communications Magazine}, vol.~50, no.~5, pp. 148--161, May 2012.

\bibitem{yousaf2017}
F.~Z. Yousaf, M.~L.~F. de Almeida, M.~Z. Shakir, M.~A. Imran, and K.~K.
Wong, ``Densification of cellular networks: A technical and economic perspective,'' \emph{IEEE Communications Magazine}, vol.~55, no.~12, pp. 170--177, Dec. 2017.

\bibitem{cisco2020}
Cisco, ``Cisco Annual Internet Report (2018--2023) White Paper,'' Mar. 9, 2020.

\bibitem{ericsson2023}
Ericsson, ``Ericsson Mobility Report November 2023,'' 2023.

\bibitem{ide2022}
C.~Ide, ``Improving mobile network performance and efficiency with machine learning: A survey,'' \emph{IEEE Communications Surveys \& Tutorials}, vol.~24, no.~1, pp. 287--316, Firstquarter 2022.

\bibitem{silva2019}
M.~D.~D. de~Silva, T.~K.~D.~M. de~Silva, and G.~M.~R.~I. Godaliyadda, ``A Survey of QoS in Mobile Cellular Networks,'' in \emph{Proc. 2019 14th Conference on Industrial and Information Systems (ICIIS)}, 2019, pp. 385--390.

\bibitem{kliks2022}
A.~Kliks, P.~Kryszkiewicz, and H.~Bogucka, ``Cost Efficiency in 5G-and-Beyond Mobile Networks: A Survey on CAPEX/OPEX Reduction and Business Model Transformation,'' \emph{IEEE Communications Surveys \& Tutorials}, vol.~24, no.~1, pp. 317--347, Firstquarter 2022.

\bibitem{gsma2023}
GSMA, ``The State of Mobile Internet Connectivity Report 2023,'' 2023.

\bibitem{minges2015}
R.~Minges, ``Exploring the Relationship Between Broadband and Economic Growth,'' World Bank, Washington, DC, 2015.

\bibitem{bastagli2018}
S.~Bastagli and I.~L.~G. Rios, ``The contribution of mobile communications to the Sustainable Development Goals,'' \emph{IEEE Communications Magazine}, vol.~56, no.~8, pp. 118--124, Aug. 2018.

\bibitem{alemaishat2021}
S.~Alemaishat \emph{et~al.}, ``Data-Driven Analytics for Future Wireless Networks: A Survey,'' \emph{IEEE Access}, vol.~9, pp. 160086--160117, 2021.

\bibitem{declaro2022}
A.~P. Declaro \emph{et~al.}, ``A machine learning-based approach in predicting cellular network traffic in the Philippines using open-source cell tower data,'' in \emph{Proc. 2022 IEEE 14th International Conference on Humanoid, Nanotechnology, Information Technology, Communication and Control, Environment, and Management (HNICEM)}, 2022, pp. 1--6.

\bibitem{tiamiyu2022}
I.~A. Tiamiyu, R.~A. Abd-Alhameed, J.~M. Noras, and S.~M.~R. Jones, ``Techno-Economic Considerations for Sunsetting 2G/3G Mobile Networks,'' in \emph{Proc. 2022 International Conference on UK-China Emerging Technologies (UCET)}, 2022, pp. 1--4.

\bibitem{koutitas2023}
G.~Koutitas, ``Challenges in the 2G/3G network sunset,'' \emph{IEEE Potentials}, vol.~42, no.~1, pp. 27--31, Jan.-Feb. 2023.

\bibitem{redana2018}
S.~Redana, A.~Cattoni, and M.~Tirkkonen, ``Challenges and solutions for 2G/3G/4G/5G multi-connectivity,'' in \emph{Proc. 2018 European Conference on Networks and Communications (EuCNC)}, 2018, pp. 1--5.

\bibitem{elbazzal2023}
Z.~El-bazzal, M.~F.~A. Abdullah, and S.~A. Ghaith, ``Predicting cellular network performance using machine learning techniques with OpenCelliD data,'' in \emph{Proc. 2023 International Conference on Advances in Network Computing and Information Security (ANCIS)}, 2023, pp. 1--6.

\bibitem{pawelczak2020}
P.~Pawełczak \emph{et~al.}, ``Is OpenCelliD a useful data source for academic research? A case study in cellular network performance,'' in \emph{Proc. 2020 IEEE 3rd 5G World Forum (5GWF)}, 2020, pp. 313--318.

\bibitem{goldberg2017}
M.~A. Goldberg and L.~A.~P.~P. da~Silva, ``A survey of reverse geocoding methods,'' in \emph{Proc. 2017 IEEE International Geoscience and Remote Sensing Symposium (IGARSS)}, 2017, pp. 3201--3204.

\bibitem{sengar2022}
N.~P.~S.~Sengar and P.~K. Sharma, ``A survey on network traffic monitoring and analysis,'' in \emph{Proc. 2022 2nd International Conference on Advance Computing and Innovative Technologies in Engineering (ICACITE)}, 2022, pp. 195--200.

\bibitem{dasilva2021}
H.~N.~D. Da~Silva, J.~J.~P.~C. Rodrigues, and R.~A.~L. Rabêlo, ``A Survey on Cellular Network Traffic Prediction,'' \emph{IEEE Access}, vol.~9, pp. 165383--165403, 2021.

\bibitem{hodge2010}
V.~J. Hodge and J.~Austin, ``A Survey of Outlier Detection Methodologies,'' \emph{IEEE Transactions on Knowledge and Data Engineering}, vol.~22, no.~8, pp. 1045--1058, Aug. 2010.

\bibitem{wu2019}
C.~F.~J. Wu, ``Challenges of statistical models and data analytics in the information age,'' in \emph{Proc. 2019 IEEE International Conference on Big Data (Big Data)}, 2019, pp. 1--1.

\bibitem{chochliouros2006}
I.~Chochliouros \emph{et~al.}, ``Challenges in the Provision of Mobile Services in Rural and Remote Areas,'' in \emph{Proc. 2006 IEEE 17th International Symposium on Personal, Indoor and Mobile Radio Communications}, 2006, pp. 1--5.

\bibitem{gupta2015}
A.~Gupta and R.~K. Jha, ``A survey of 5G network: Architecture and emerging technologies,'' \emph{IEEE Access}, vol.~3, pp. 1206--1232, 2015.

\bibitem{khan2020dimensioning}
M.~U.~Khan, A.~Garc{\'i}a-Armada, and J.~J.~Escudero-Garz{\'a}s,
``Service-based network dimensioning for 5G networks assisted by real data,''
\emph{IEEE Access}, vol.~8, pp.~129193--129212, 2020.

\bibitem{khan2020heuristic}
M.~U.~Khan, M.~Azizi, A.~G.~Armada, and J.~J.~Escudero Garz{\'a}s,
``Heuristic for network planning based on 5G services,''
in \emph{Advances in Smart Technologies Applications and Case Studies},
A.~El~Moussati, K.~Kpalma, M.~G.~Belkasmi, M.~Saber, and S.~Gu{\'e}gan, Eds.
Cham: Springer International Publishing, 2020, pp.~3--14.

\bibitem{khan2020clustering}
M.~U.~Khan, M.~Azizi, A.~Garc{\'i}a-Armada, and J.~J.~Escudero-Garz{\'a}s,
``Unsupervised clustering for 5G network planning assisted by real data,''
2020.

\end{thebibliography}
\end{document}